\documentclass[aps,pra,twocolumn,groupedaddress,amsmath,amssymb]{revtex4}

\usepackage{bm}
\usepackage{graphicx}
\newcommand{\beq}{\begin{equation}}
\newcommand{\eeq}{\end{equation}}
\newcommand{\lsup}[2]{\mbox{$^{#1}{#2}$}}

\makeatletter
\def\gsim{\compoundrel>\over\sim}
\def\lsim{\compoundrel<\over\sim}
\def\compoundrel#1\over#2{\mathpalette\compoundreL{{#1}\over{#2}}}
\def\compoundreL#1#2{\compoundREL#1#2}
\def\compoundREL#1#2\over#3{\mathrel
   {\vcenter{\hbox{$\m@th\buildrel{#1#2}\over{#1#3}$}}}}
\makeatother

\newcommand{\MOT}{\textsc{mot}}
\newcommand{\MCP}{\textsc{mcp}}

\newcommand{\CCD}{\textsc{ccd}}
\newcommand{\AOM}{\textsc{aom}}
\newcommand{\PI}{\textsc{pi}}
\newcommand{\AI}{\textsc{ai}}

\newcommand{\onemu}{\mbox{$\mspace{1.0mu}$}}
\newcommand{\micro}{\mu}

\begin{document}


\title{Homonuclear ionizing collisions of laser-cooled metastable helium atoms}


\author{R.\ J.\ W.\ Stas}
\author{J.\ M.\ McNamara}
\author{W.\ Hogervorst}
\author{W.\ Vassen}
\affiliation{Laser Centre Vrije Universiteit, De Boelelaan 1081, 1081 HV Amsterdam, The Netherlands}


\date{\today}

\begin{abstract}
We present a theoretical and experimental investigation of
homonuclear ionizing collisions of laser-cooled metastable
($2~\lsup{3}{\text{S}}_1$) helium atoms, considering both the 
fermionic
$^3$He and 
bosonic
$^4$He isotope. The theoretical description combines quantum
threshold behavior, Wigner's spin-conservation rule and quantum
statistical symmetry requirements in a single-channel model,
complementing a more complete close-coupling theory that has been
reported for collisions of metastable $^4$He atoms. The model is
supported with measurements (in the absence of light fields) of
ionization rates in magneto-optically trapped samples, that contain
about $3 \times 10^8$ atoms of a single isotope. The ionization
rates are determined from measurements of trap loss due to
light-assisted collisions combined with comparative measurements of
the ion production rate in the absence and presence of trapping
light. Theory and experiment show good agreement.
\end{abstract}


\maketitle


\section{\label{sec:int}Introduction}

Soon after the first demonstration of laser cooling of neutral
atoms, it was recognized that collisions between atoms have a
profound effect on the physics of laser-cooled atomic gases
\cite{goul88}. Light-assisted collisions and other inelastic
collision processes constitute loss channels that limit the
attainable atomic densities in laser-cooled samples. The
magneto-optical trap (\MOT) was found to be a versatile tool for
accurate investigation of these inelastic collisions \cite{pren88},
that occur at collision energies of $10^{-7}$~eV in samples with
temperatures around 1~mK. Measuring loss rates from the trap, the
cross section for inelastic collisions can be determined with great
precision \cite{wein03}. Experiments as well as theoretical work,
almost exclusively carried out in alkali systems, have provided a
wealth of information on cold collisions \cite{wein03}, that play a
critical role in many research areas that have emerged since the
advent of laser cooling \cite{burn02}.

Laser-cooled helium atoms in the metastable $2~\lsup{3}{\text{S}}_1$
state, denoted He*, provide unique opportunities to study cold
ionizing collisions,
\begin{alignat}{2}
\text{He*}+\text{He*} & \rightarrow \text{He} + \text{He}^+ + e^- & \quad &  \text{(\PI)} \label{eq3:PI}, \\
\text{He*}+\text{He*} & \rightarrow \text{He}_2^+ + e^- && \text{(\AI)} \label{eq3:AI},
\end{alignat}
where \PI\ stands for Penning ionization and \AI\ for associative
ionization. The simple structure of the helium atom allows for
theoretical exercises that are relatively uncomplicated, while
experiments profit by the possibility of direct detection of ions
with charged-particle detectors. Furthermore, large numbers of atoms
($\sim 3 \times 10^8$), either $^3$He* or $^4$He*, can be confined
in a \MOT\ \cite{stas04}, so that differences between collisions of
fermionic ($^3$He*) and bosonic atoms ($^4$He*) can be investigated.

In this article, a theoretical and experimental investigation of
homonuclear ionizing collisions of $^3$He* and $^4$He* atoms is
presented, with particular regard to isotopic differences in
ionizing collisions. The theoretical model is a single-channel
calculation of the ionization cross section and rate coefficient for
ionizing collisions, that can be applied to both isotopes. In the
experiments, magneto-optically trapped samples of $^3$He* or $^4$He*
atoms are investigated and ionization rates are measured with a
micro-channel plate (\MCP) detector. The measurements are performed
in the absence of trapping light, as optical excitation by
near-resonant light alters the dynamics of cold collisions
\cite{suom96,wein03}, thereby strongly enhancing the ionization rate
\cite{bard92,mast98,mast98a,tol99,kuma99,kuma00,brow00,pere01a}.

Combining theory and experiment, the present work extends
previous studies and may be used to unravel inconsistencies in
reported experimental results \cite{leo01}. The transparent
theoretical model complements close-coupling calculations that have
been performed for collisions of $^4$He* atoms
\cite{vent99,vent00,leo01}. Other studies of $^4$He* collisions
provide little detail \cite{juli89,mast98} or focus exclusively on
collisions of spin-polarized atoms (to investigate the feasibility
of Bose-Einstein condensation) \cite{shly94,fedi96}, while another
theoretical study, that applies to both $^3$He* and $^4$He*
collisions \cite{kuma99}, is based on questionable assumptions. The
extensive description of the experimental procedure and exhaustive
explanation of the data analysis procedure provides clear insight
into the origin of experimental ionization rate coefficients
and can be used to evaluate other experimental studies of ionizing
collisions of He* atoms \cite{mast98,kuma99,tol99,hers00}.

The article is organized as follows. Section~\ref{sec:th} presents
the theoretical model for homonuclear ionizing collisions of
laser-cooled helium isotopes. First a simplified expression for the
ionization cross section is derived (Sec.~\ref{sec:th:cross}). Using
an effective molecular potential (Sec.~\ref{sec:th:pot}), partial
wave ionization cross sections are derived from numerical solutions
of the one-dimensional Schr\"odinger equation
(Sec.~\ref{sec:th:num}) and the corresponding partial wave
ionization rate coefficients are calculated
(Sec.~\ref{sec:th:coef}). Quantum statistical symmetry requirements
are taken into account (Sec.~\ref{sec:th:symm}) to derive the rate
coefficient for samples of either $^3$He* or $^4$He* atoms with a given
population of magnetic sublevels. Finally, the results are compared
with other theoretical results (Sec.~\ref{sec:th:unpol}).
Section~\ref{sec:exp} describes the measurement of the ionization
rate coefficients in laser-cooled samples of $^3$He* or $^4$He*
atoms. Subdividing loss mechanisms into ionizing and non-ionizing
contributions, and distinguishing between linear and quadratic trap
loss, an overview of trap loss mechanisms occurring in our He*
samples is presented and estimates of ionization and trap loss rates
are derived (Sec.~\ref{sec:exp:loss}). It is shown that ionization
rates can be deduced from trap loss measurements if the contribution
of ionizing mechanisms to trap loss is determined. Measurements of
trap loss due to light-assisted collisions are presented
(Sec.~\ref{sec:exp:light}). Comparative ion production rate
measurements in the absence and presence of trapping light are used
to determine the ionization rate coefficients in the absence of
light fields (Sec.~\ref{sec:exp:dark}). Finally, the results are
compared with our theoretical predictions (Sec.~\ref{sec:exp:comp}).

Section~\ref{sec:disc} presents a discussion of the theoretical and
experimental results, as well as conclusions and prospects.

\section{\label{sec:th}Theoretical model}

The ionizing collisions of Eqs.~(\ref{eq3:PI}) and (\ref{eq3:AI})
are highly exothermic, as the internal energy of two He* atoms
exceeds the 24.6~eV ionization energy of the He atom by more than
15~eV. As differences between \PI\ and \AI\ are unimportant for the
work presented here (the reaction mechanisms have been discussed in
detail in Ref.~\cite{yenc84}), we will not distinguish between them and use
the term \PI\ to denote both processes.

The interaction that drives the auto-ionizing transitions of
Eqs.~(\ref{eq3:PI}) and (\ref{eq3:AI}) is of an electrostatic nature
\cite{yenc84}, so that it only induces transitions between molecular
states of equal total electronic spin. Therefore, ionization rates
associated with the reactions of Eqs.~(\ref{eq3:PI}) and
(\ref{eq3:AI}) depend on the total spin states on the reactant and
product side of the reaction formulas. For both reactions, the
reactants carry an electronic spin of $s=1$ and can form total spin
states with $S=0,1$ or 2, while the products, carrying
$s=\frac{1}{2}$ (except for ground state helium, which carries no
electronic spin), can only form states with $S=0$ or 1. Clearly,
total electronic spin can only be conserved if $S=0$ or 1, and a
\PI\ reaction with $S=2$ would involve a violation of spin
conservation.
It has been shown \cite{shly94,fedi96} that a very weak spin-dipole
magnetic interaction can induce spin flips and mediate \PI\ in
collisions of He* atoms with $S=2$. The corresponding ionization
rate is four orders of magnitude smaller than those of
collisions with $S=0$ and $S=1$, for which total electronic spin is
conserved \cite{sirj02}. The strong suppression of \PI\ by spin
conservation is known as Wigner's spin-conservation rule
\cite{mass71} and has been observed for He* collisions in a gas
discharge \cite{hill72}, a laser-cooled sample \cite{hers00} and a
Bose-Einstein condensate \cite{robe01,pere01,tych05a}. In heavier
metastable rare-gas systems, Wigner's spin-conservation rule does
not apply \cite{doer98}, and polarized samples can be used for the
investigation of quantum statistical effects in ionizing collisions
\cite{kato95,orze99}.

At the mK temperatures of a laser-cooled sample of He* atoms,
collisions occur at relative kinetic energies $E = \mu v_\text{r}^2
/2 \approx 10^{-7}$~eV, where $\mu=m/2$ is the reduced mass of the
colliding atoms (with $m$ the mass of the He atom) and $v_\text{r}$
is the relative velocity between the atoms. As the de Broglie
wavelength of atomic motion $\Lambda=h/\mu v_\text{r} \approx 250 \,
a_0$ (with $h$ Planck's constant) is much larger than the typical
scale of the interatomic potential, collisions are dominated by
quantum threshold behavior
\cite{beth35,wign48,juli89,juli93,mies00}. In this case, the
collision process can be described conveniently using the partial
wave method \cite{cohe77}: the ionization cross section, written
as a sum of partial wave contributions,
\beq
\sigma^\text{(ion)} = \sum_\ell \sigma_\ell^\text{(ion)},
\label{eq:pwexpansion}
\eeq
is dominated by only a few partial waves $\ell$. For inelastic
exothermic collisions, such as Penning ionizing collisions, the
quantum threshold behavior of the $\ell$th partial cross section is
given by \cite{juli89,juli93,wein03}
\beq
\sigma_\ell^\text{(ion)} \! \propto k^{2\ell-1} \quad \text{if $k \rightarrow 0$.}
\label{eq3:quantthresh}
\eeq
Here, $k = (2 \mu E/\hbar^2)^{1/2}$ is the wave vector of the
asymptotic relative motion of the colliding atoms, with $\hbar=h/2
\pi$. In a sufficiently cold sample of He* atoms, the cross section
for Penning ionizing collisions is dominated by the s-wave
contribution $\sigma_0^\text{(ion)}$, which diverges as $1/k$ if $k
\rightarrow 0$ \cite{beth35}. Elastic collisions have very different
threshold properties: the cross section $\sigma_0^\text{(elas)}$
approaches a nonvanishing constant, $\sigma_1^\text{(elas)} \!
\propto k^{4}$ and $\sigma_{\ell>1}^\text{(elas)} \! \propto k^{6}$,
if $k \rightarrow 0$ \cite{dali99,wein03}.

For collisions of He* atoms, partial wave cross section
$\sigma_\ell^\text{(ion)}$ can be derived from the solution of an
effective one-dimensional potential scattering problem
(Secs.~\ref{sec:th:cross}--\ref{sec:th:coef}). Restrictions
imposed by the symmetry postulate on the partial waves that
contribute to the cross section of Eq.~(\ref{eq:pwexpansion}) can be
taken into account thereafter (Sec.~\ref{sec:th:symm}).

\subsection{\label{sec:th:cross}Ionization cross section}
From a semi-classical point of view, two events can be
distinguished in the process of a cold ionizing collision of two He*
atoms: (1) elastic scattering of the atoms by the interaction
potential $V(R)$, with $R$ the internuclear distance, and (2)
Penning ionization that occurs when the two electron clouds start to
overlap \cite{yenc84}. As collision energies are small, the elastic scattering
occurs at a relatively large internuclear distance $R \gsim 100 \,
a_0$. For partial waves $\ell
> 0$, the radial wave function $u_{k\ell}(R)$ (cf.
Eq.~(\ref{eq3:radialwaveeq})) is scattered by the centrifugal
barriers, while scattering for $\ell=0$ takes places at the
internuclear distance where the local de Broglie wavelength
$\Lambda(R) = h/\{2 \mu [E-V(R)]\}^{1/2}$ becomes comparable to the
size of the potential, i.e. $\text{d}\Lambda(R)/\text{d}R \approx 1$
\cite{juli89,juli93}. As the electron clouds of both atoms start to
overlap at small internuclear distance $R \approx 5 \, a_0$
\cite{mull91}, the elastic scattering process can be considered to
precede the inelastic process of ionization.

In the spirit of Ref.~\cite{orze99}, we assume that the two subsequent
processes can be treated separately. As \PI\ is a
strong inelastic exothermic process, we can write the ionization
cross section for collisions with total electronic spin $S$ as
\cite{orze99}
\beq
\lsup{(2S+1)}{\sigma}^\text{(ion)} = \frac {\pi}{k^2} \sum_\ell (2\ell+1) \: \lsup{(2S+1)}{P}_\ell^\text{(tun)} \: \lsup{(2S+1)}{P}^\text{(ion)},
\label{eq3:sigma_ion}
\eeq
where $\lsup{(2S+1)}{P}_\ell^\text{(tun)}$ is the probability for
the atoms to reach a small internuclear distance, and
$\lsup{(2S+1)}{P}^\text{(ion)}$ is the probability for ionization to
occur at that place. As total spin $S$ is strongly conserved during
ionization, $\lsup{(2S+1)}{P}^\text{(ion)}$ is very small ($\ll 1$)
for collisions that violate Wigner's spin-conservation rule. Here,
we neglect ionizing collisions with $S=2$ and assume that
$\lsup{5}{P}^\text{(ion)}=0$. As ionization occurs with essentially
unit probability for the other spin states (M\"uller \emph{et al.\
}\cite{mull91} report an ionization probability of 0.975), we set
$\lsup{1}{P}^\text{(ion)}=\lsup{3}{P}^\text{(ion)}=1$.

The calculation of cross sections
$\lsup{(2S+1)}{\sigma}^\text{(ion)}$ is reduced to the determination
of partial wave tunneling probabilities
$\lsup{(2S+1)}{P}_\ell^\text{(tun)}$. The energy dependence of the
probabilities $\lsup{(2S+1)}{P}_\ell^\text{(tun)}$ gives rise to an
energy dependent $\lsup{(2S+1)}{\sigma}^\text{(ion)}$, that displays
the quantum threshold behavior of the inelastic collisions of
Eq.~(\ref{eq3:quantthresh}). To calculate
$\lsup{(2S+1)}{P}_\ell^\text{(tun)}$, we need to consider the
interaction potential of the colliding atoms.

\subsection{\label{sec:th:pot}Effective potential}
\begin{figure}[t]
\begin{center}
\includegraphics[width=\columnwidth,keepaspectratio]{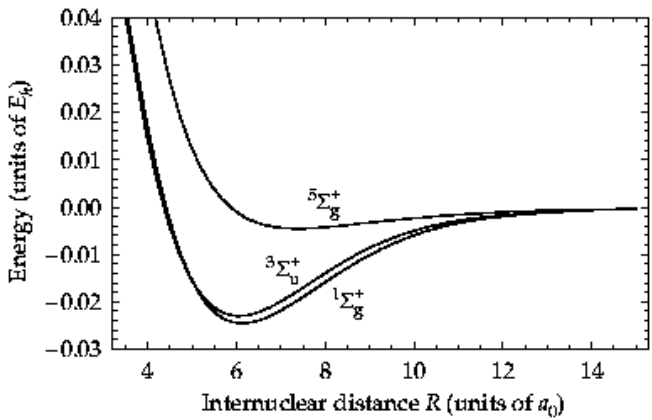}
\caption{\label{fig3:shortrangepot}Ab initio potential energy curves in atomic units, calculated
for $^4$He* by M\"uller \emph{et al.~}\cite{mull91}.}
\end{center}
\end{figure}
As the helium atom, with only two electrons and a nucleus, has a
relatively simple structure, interatomic potentials can be
calculated with high accuracy. Figure \ref{fig3:shortrangepot} shows
the short-range part ($3.5 \, a_0 < R < 14.0 \, a_0$) of the
potential curves for two metastable $2~\lsup{3}{\text{S}}_1$ helium
atoms obtained from M\"uller \emph{et al.~}\cite{mull91}. The curves
are the result of ab initio calculations in the Born-Oppenheimer
approximation, where total electronic spin $S$ is a good quantum
number. The possible values $S=0$, $S=1$ and $S=2$ correspond to a
singlet, triplet and quintet potential ($^1V(R)$, $^3V(R)$ and
$^5V(R)$), respectively. In Fig.~\ref{fig3:shortrangepot}, the
curves are labeled in Hund's case (a) notation $^{2S+1} \Lambda
_\text{g/u}^\pm$, where $\Lambda = |M_L|$ with $M_L$ the quantum
number of the projection of the total electronic orbital angular
momentum onto the internuclear axis of the molecule, g/u stands for
gerade or ungerade, i.e.\ positive or negative symmetry under
inversion of all electronic coordinates of the molecule, and $\pm$
indicates positive or negative symmetry under reflection through a
plane including the internuclear axis \cite{land65}. As both
electrons of the He* atom are in s-states, colliding atom pairs can
only have zero total orbital angular momentum, indicated by $\Lambda
= \Sigma$.

The potentials can be extended to large internuclear distance using
a calculation of the dispersion interaction of two He* atoms;
for a multipole expansion
$-C_6/R^6 - C_8/R^8 - C_{10}/R^{10}$, 
dispersion coefficients $C_6=3276.680$~a.u., $C_8=210566.55$~a.u.\
and $C_{10}=21786760$~a.u. have been reported ~\cite{yan98,spel93}. 
Here, we construct potentials valid for
$R>3.5\, a_0$ by fitting the short-range potential curves of
Fig.~\ref{fig3:shortrangepot} smoothly onto the long-range
dispersion interaction around $20 \, a_0$ by interpolation
using a cubic spline fitted to $R^6 \times \lsup{(2S+1)}{V(R)}$.

The elastic scattering for collisions with $S=0$ and $S=1$ is
governed by potentials $^1V(R)$ and $^3V(R)$, respectively. Within
the framework of the partial wave method, potential scattering by
$\lsup{(2S+1)}{V}(R)$ is described by the radial wave equation
\cite{joac75}
\begin{multline}
-\frac{\hbar^2}{2 \mu} \frac{\text{d}^2}{\text{d}R^2} u_{k\ell}(R) \\ +
 \left[\frac{\hbar^2 \ell(\ell+1)}{2 \mu R^2}+\lsup{(2S+1)}{V}(R)-E \right]
 u_{k\ell}(R) = 0,
\label{eq3:radialwaveeq}
\end{multline}
where $\ell$ is the quantum number of the relative angular momentum
and $u_{k\ell}(r)$ is the radial wave function.
Equation~(\ref{eq3:radialwaveeq}) can be interpreted as a
one-dimensional Schr\"odinger equation ($R \geq 3.5 \, a_0$),
describing the potential scattering of a particle of mass $\mu$ by
effective potential
\beq
\lsup{(2S+1)}{V}_\ell(R) = \lsup{(2S+1)}{V}(R)+ \frac{\hbar^2 \ell(\ell+1)}{2 \mu R^2},
\eeq
where the second term is the well-known centrifugal potential.

To calculate the probability $\lsup{(2S+1)}{P}_\ell^\text{(tun)}$ of
atom pairs to reach the distance where ionization occurs, we modify
the effective potential curves to simulate the ionization process.
We set the curves to a constant value for small internuclear
distances, and extend the range of $R$ to negative values
\cite{orze99},
\beq
\lsup{(2S+1)}{\widetilde{V}}_\ell(R) = \begin{cases}
\lsup{(2S+1)}{V}_\ell(R_0)& R \leq R_0, \\
\lsup{(2S+1)}{V}_\ell(R) & R > R_0, \end{cases}
\eeq
where $R_0 = 6.1 \, a_0$ is chosen to be the location of the
potential curve minimum. In this way, we avoid reflections of the
radial wave function from artificial features of the potential
energy curve at $R_0$. Modeling the interatomic interaction by
$\widetilde{V}_\ell(R)$, potential scattering is described by the
one-dimensional Schr\"odinger equation ($-\infty < R < \infty$)
\beq
-\frac{\hbar^2}{2 \mu} \frac{\text{d}^2}{\text{d}R^2} u_{k\ell}(R) +
 \left[\lsup{(2S+1)}{\widetilde{V}}_\ell(R)-E \right]
 u_{k\ell}(R) = 0,
\label{eq3:1dwaveeq}
\eeq
and ionizing collisions correspond to the transmission of the
relative particle to $R<0$: for atoms that reach the region of small
$R$, where ionization takes place, the corresponding relative
particle will propagate freely to $R = -\infty$ and never reflect
back to $R>R_0$. The disappearance of the particle to $R<0$ results
in a loss of probability flux.

\begin{figure}[t]
\begin{center}
\includegraphics[width=\columnwidth,keepaspectratio]{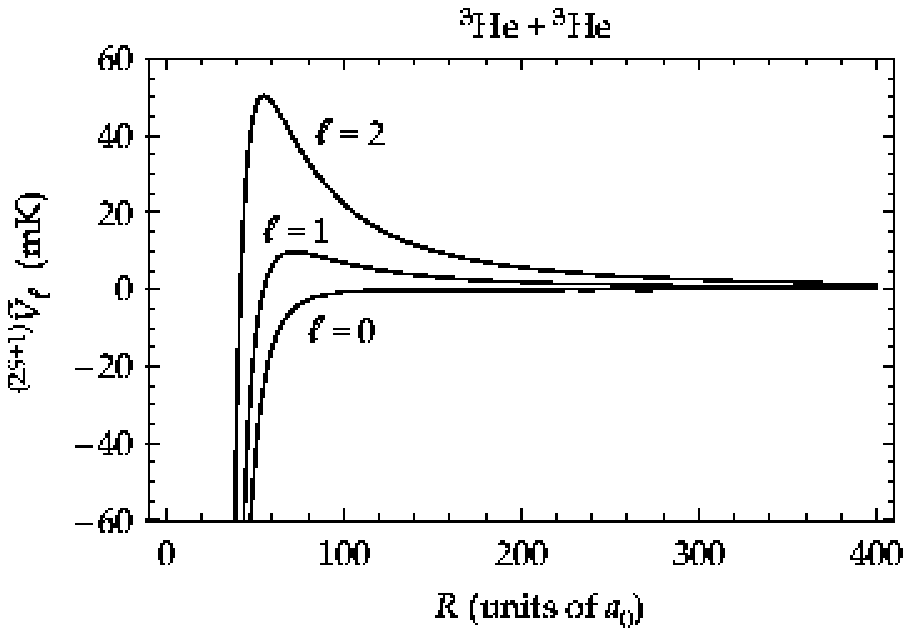}
\includegraphics[width=\columnwidth,keepaspectratio]{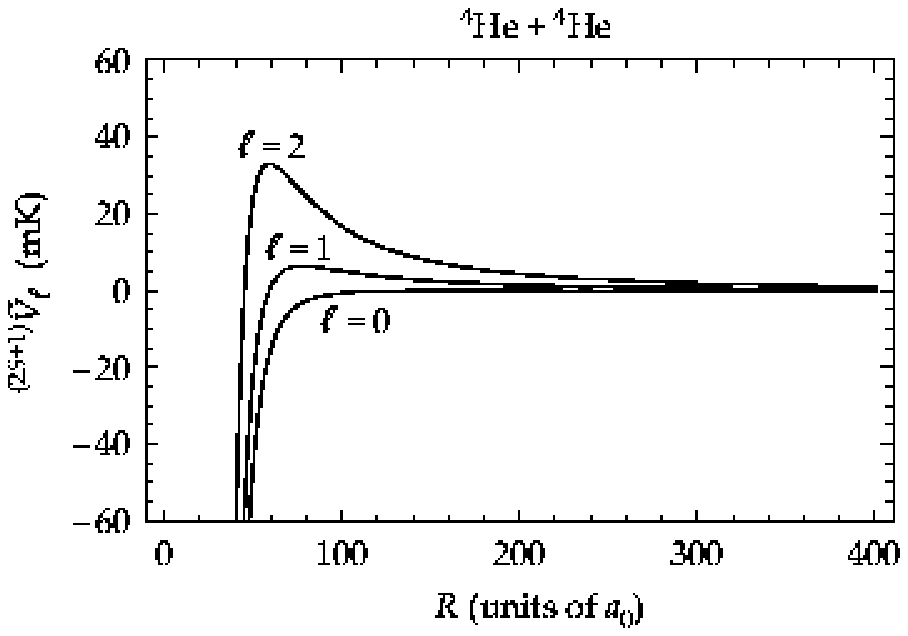}
\caption{\label{fig3:rotbar}Potentials $\lsup{(2S+1)}{\widetilde{V}}_\ell(R)$ with centrifugal barriers. The
centrifugal interaction is smaller than the short-range attraction by five orders of magnitude.}
\end{center}
\end{figure}
Figure~\ref{fig3:rotbar} shows plots of $\widetilde{V}_0(R)$,
$\widetilde{V}_1(R)$, and $\widetilde{V}_2(R)$ for homonuclear
collisions of both $^3$He* and $^4$He* atoms, where we have used the
atomic mass $m=3.01603$~u for $^3$He* and $m=4.0026$~u for $^4$He*,
with $1 \: \text{u} = 1822.89$~a.u. For an atomic sample with a
thermal velocity distribution with temperature $T$, the mean
collision energy is given by $\langle E \rangle =\frac{3}{2}
k_\text{B} T$; this relation is used to express the potentials in
units of temperature.

The barriers formed by centrifugal potentials with $\ell=1$ and 2
are five orders of magnitude smaller than the potential energy
associated with the short-range attraction of the colliding atoms.
However, even the lowest barrier, with $\ell=1$, is large compared
to the temperature of 1~mK that is typical of samples of
laser-cooled He* atoms, and the barrier heights increase with
increasing $\ell$. Therefore, probability
$\lsup{(2S+1)}{P}_\ell^\text{(tun)}$ is small for $\ell=1$ and
decreases rapidly for larger values of $\ell$.

\subsection{\label{sec:th:num}Effective potential scattering problem}
The probability of transmission through a barrier in a potential can
be calculated from the stationary states associated with the potential
\cite{cohe77}. To calculate the probability for transmission to
$R<0$ in potential $\lsup{(2S+1)}{\widetilde{V}}_\ell(R)$, we
consider stationary states that correspond to the sum of an incident
and a reflected wave for $R \gg 1$, and a single transmitted wave
for $R \leq R_0$. The transmission probability
$\lsup{(2S+1)}{P}_\ell^\text{(tun)}$ can be written as \cite{cohe77}
\beq
\lsup{(2S+1)}{P}_\ell^\text{(tun)}=\frac{J_{\,\text{tr}}}{J_{\,\text{in}}},
\eeq
where $J_{\,\text{in}}$ and $J_{\,\text{tr}}$ are the probability
fluxes associated with incident and transmitted plane waves,
respectively. Numerical methods are used to calculate
$\lsup{(2S+1)}{P}_\ell^\text{(tun)}$, for $\ell=0,1,2$ and a range of
collision energies $E$.

For a typical collision energy $E = 9.5 \times 10^{-9}$~a.u.\
($E/\frac{3}{2}k_\text{B}=2.0$~mK), the transmission probabilities
are $\lsup{1}{P}_0^\text{(tun)}=0.66$,
$\lsup{1}{P}_1^\text{(tun)}=0.086$ and
$\lsup{1}{P}_2^\text{(tun)}=5.8 \times 10^{-4}$. Clearly, reflection
is almost complete in the case of d-wave scattering (as $E$ is much
smaller than the barrier height), while transmission is significant
for p-wave scattering. In the case of s-wave scattering, there is
considerable reflection, although a centrifugal barrier is absent.
Here, quantum reflection occurs due to the mismatch between the long
asymptotic de Broglie wave and the rapidly oscillating wave in the
region of small internuclear separation. We checked that the
resulting partial wave cross sections satisfy the quantum threshold
behavior of Eq.~(\ref{eq3:quantthresh}).

To determine the dependence of the cross sections on the adapted
short-range part of the potential, we have calculated the variation
in the cross sections as a function of $R_0$ for various collision
energies. If $R_0$ is close to the location of the potential curve
minimum at $6.1\, a_0$, the variations are smallest (less than
0.2\%). Furthermore, the difference between probabilities
$\lsup{1}{P}_\ell^\text{(tun)}$ and $\lsup{3}{P}_\ell^\text{(tun)}$
at a given collision energy $E$ is only a few percent, as potentials
$\lsup{1}{\widetilde{V}}(R)$ and $\lsup{3}{\widetilde{V}}(R)$ differ
very little in the region where elastic scattering takes place:
$|\lsup{3}{\widetilde{V}}(R)-\lsup{1}{\widetilde{V}}(R)|/|\lsup{1}{\widetilde{V}}(R)|<10^{-4}$
for $R>20\, a_0$.

It has been shown in calculations that the ionization cross sections
for $^4$He* are enhanced if the s-wave scattering length
$\lsup{5}{a}$ associated with the quintet potential is near a
singularity \cite{leo01}. The s-wave scattering length describes
elastic collisions in the low-temperature limit \cite{dali99} and
shows a singularity (goes through $\pm \infty$) whenever a bound
state is removed from the potential \cite{leo01}. From experiments,
it has been determined that 
$\lsup{5}{a}= 7.6$~nm with
an error of 0.6 nm \cite{kim05}.
The scattering length is sufficiently far from the singularity to
neglect enhancement of the ionization cross sections. For $^3$He*
atoms, the s-wave scattering length for $S=2$ is predicted to be 
$5.0\text{~nm}<\lsup{5}{a}<6.0$~nm \cite{dick04}. This is also
sufficiently small to neglect effects on the ionization cross
sections.

\subsection{\label{sec:th:coef}Ionization rate coefficient}
The ion production rate $\text{d}N_\text{ion}/\text{d}t$ in a
magneto-optically trapped atomic sample of $^3$He* or $^4$He* atoms
can be expressed in terms of an ionization rate coefficient
$K$ (particle$^{-1}$ cm$^3$/s),
\beq
\frac{\text{d}N_\text{ion}(t)}{\text{d}t} = K \iiint n^2(\bm{r},t)
\, \text{d}^3r. \label{eq:ionrate}
\eeq
The rate coefficient depends on the temperature $T$ of the sample and
can be written as \cite{juli89}
\beq
K(T) = \int_0^\infty \sigma^\text{(ion)}\!(v_\text{r}) \;\;
P_T^\text{(MB)}\!(v_\text{r}) \;\; v_\text{r}\, dv_\text{r},
\eeq
with
\beq
P_T^\text{(MB)}\!(v_\text{r}) = \sqrt{\frac{2}{\pi}} \; \frac{v_\text{r}^2}{(k_\text{B} T /
\mu)^{3/2}} \; \exp \!\left(-\frac{v_\text{r}^2}{k_\text{B} T / \mu} \right),
\eeq
the Maxwell-Boltzmann distribution for the relative velocity in the
atomic sample under study, with $\langle v_\text{r}^2 \rangle^{1/2}
= (k_\text{B} T / \mu)^{1/2}=(2 k_\text{B} T / m)^{1/2}$.

\begin{figure}[t]
\begin{center}
\includegraphics[width=\columnwidth,keepaspectratio]{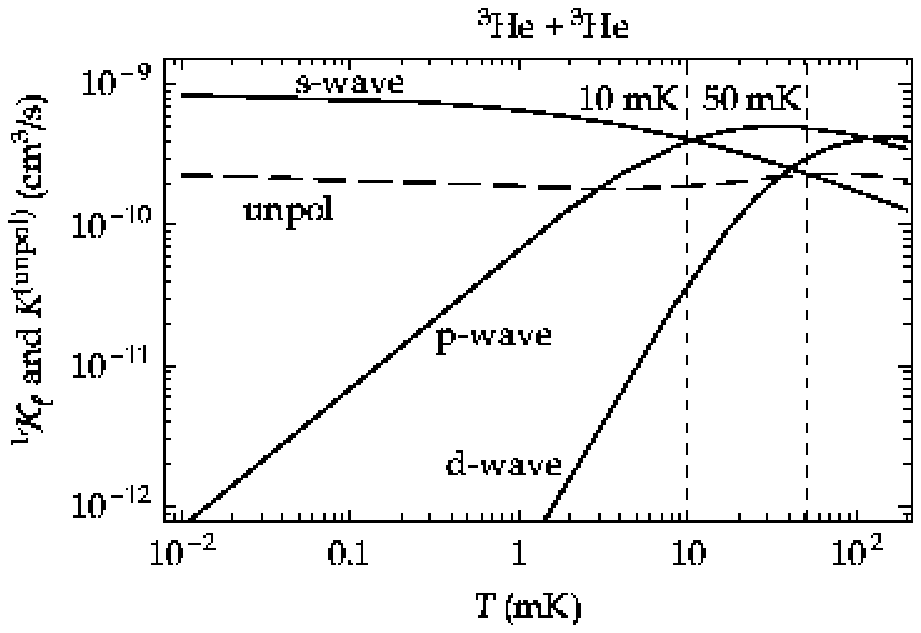}
\includegraphics[width=\columnwidth,keepaspectratio]{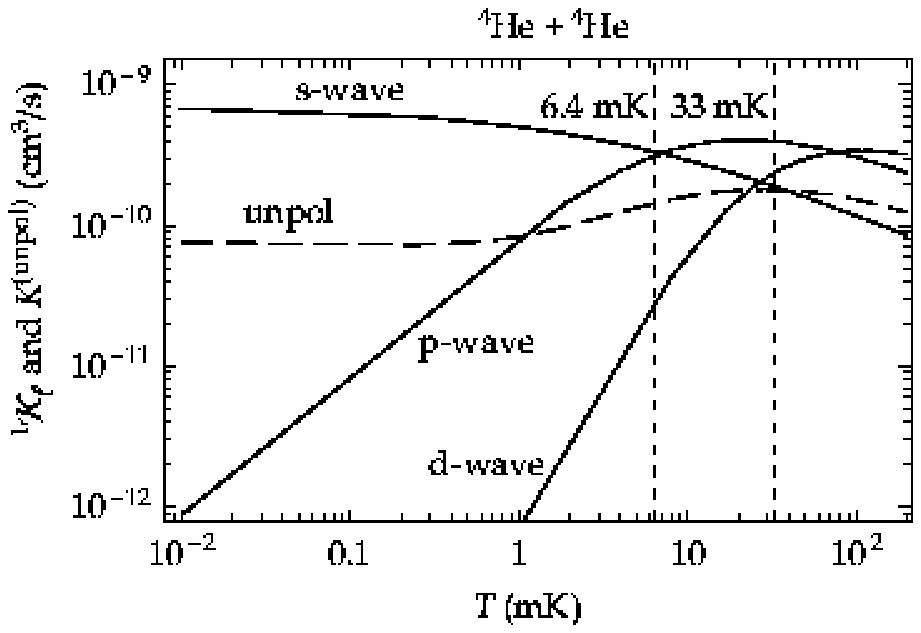}
\caption{\label{fig3:bunpol}Partial wave ionization rate
coefficients $\lsup{1}{\cal K}_\ell$ and unpolarized rate
coefficient $K^\text{(unpol)}$ for $^3$He* and $^4$He*
($\lsup{3}{\cal K}_\ell$ coefficients are, within a percent, equal to corresponding $\lsup{1}{\cal K}_\ell$ coefficients).}
\end{center}
\end{figure}
Correspondingly, we can define partial wave ionization rate
coefficients,
\beq
\lsup{(2S+1)}{{\cal K}}_\ell(T) = \int_0^\infty \lsup{(2S+1)}{\sigma}^\text{(ion)}_\ell\!(v_\text{r}) \;\;
P_T^\text{(MB)}\!(v_\text{r}) \;\; v_\text{r}\, dv_\text{r},
\eeq
and the ionization rate coefficient associated with potential
$\lsup{(2S+1)}V(R)$,
\beq
\lsup{(2S+1)}{K}(T) = \sum_\ell \lsup{(2S+1)}{{\cal K}}_\ell(T).
\label{eq3:Ksum}
\eeq
Figure~\ref{fig3:bunpol} shows plots of the rate coefficients
(calculated with numerical integration) for a temperature range from
10~$\micro$K to 100~mK, for collisions of $^3$He*
or $^4$He* atoms. The
energy of the centrifugal barriers for $\ell=1$ and $\ell=2$ is
indicated by dashed vertical lines. For temperatures $T<5$~mK, the
contribution of the d-wave becomes very small and can be ignored if
an accuracy of 5\% is sufficient.

The quantum threshold behavior of the rate coefficients is given by
\beq
\lsup{(2S+1)}{{\cal K}}_\ell \propto k^{2\ell} \quad \text{if $k \rightarrow 0$}.
\eeq
In particular, we find $\lsup{(2S+1)}{{\cal K}}_0 \rightarrow
\text{constant}$ if $k \rightarrow 0$, as the divergence of
$\sigma_0$ is canceled by $v= \hbar k / \mu$. The differences
between the two isotopes are less than 50\%.

We have neglected the atomic hyperfine structure of the interatomic
potentials for $^3$He*. As it is four orders of magnitude smaller
than the attractive interaction at short range, its effect on
$\lsup{(2S+1)}{\sigma}_\ell^\text{(ion)}$ and $\lsup{(2S+1)}{{\cal
K}}_\ell$ is negligible.

\subsection{\label{sec:th:symm}Symmetrization of scattering states}

Although the interatomic interaction is almost identical in the case
of $^3$He* and $^4$He*, giving rise to partial wave contributions
$\lsup{(2S+1)}{\sigma}_\ell^\text{(ion)}$ and $\lsup{(2S+1)}{{\cal
K}}_\ell$ that are similar, the composition of the total ionization
cross section or rate coefficient from these contributions is very
different for the bosonic ($^4$He*) and fermionic isotope ($^3$He*).
The symmetrization postulate requires that a scattering state
describing a colliding pair of identical bosons has even symmetry
under exchange of the atoms, while a state describing identical
fermions has odd symmetry \cite{mess99,ball98}. As a result, partial
waves with improper symmetry do not contribute to the total cross
section or rate coefficient, and are excluded from the summations in
Eqs.~(\ref{eq:pwexpansion}) and~(\ref{eq3:Ksum}).

\subsubsection{Bosonic $^4\!$He*: Symmetric states}
The total electronic spin $\mathbf{S}$ and relative angular momentum
$\bm{\ell}$ of two colliding $^4$He* atoms are, to good
approximation, conserved separately \cite{shly94,fedi96}, and $S$
and $\ell$ can be considered good quantum numbers. Ignoring the
radial part of the quantum states, a basis for atom pairs is
given by
\beq
\left\{ \rule[-2mm]{0mm}{4mm} |(s_1)_A (s_2)_B , S M_S , \ell m_\ell \rangle \right\}.
\eeq
For the moment, the atoms are assumed to be distinguishable and are
labeled with $A$ and $B$. The atoms carry spins $s_1=1$ and $s_2=1$,
respectively, and the total electronic spin $S$ is the result of the
addition of $s_1$ and $s_2$. Quantum numbers $M_S$ and $m_\ell$ are
associated with the projection onto the internuclear axis of the
total electronic spin and the relative angular momentum $\ell$,
respectively.

For a system of identical bosons, such as a pair of $^4$He* atoms,
physical states are symmetric under exchange of the two atoms. Such
states are obtained applying symmetrizer $\text{S}=(1 +
\text{P}_{12})/\sqrt{2}$ to the basis vectors \cite{cohe77,burk99t},
\begin{multline}
|s_1 s_2, S M_S, \ell m_\ell \rangle \\ = \frac{1}{\sqrt{2}} \left[1 +
(-1)^{S+\ell}\right] \: |(s_1)_A(s_2)_B , S M_S , \ell m_\ell \rangle,
\end{multline}
and normalizing the result if necessary. As the states differ from
zero only if $S+\ell$ is even, we can conclude that it is not
possible to construct states with the proper symmetry if $S=0$ or
$S=2$ and $\ell$ is odd, or if $S=1$ and $\ell$ is even.
Consequently, the corresponding partial wave ionization rate
coefficients are excluded from the summation of Eq.~(\ref{eq3:Ksum})
and we can write
\begin{align}
\lsup{1}{K} &= \displaystyle\sum_\text{$\ell$ even} \lsup{1}{{\cal K}}_\ell, \label{eq3:KHe4_1} \\
\lsup{3}{K} &= \displaystyle\sum_\text{$\ell$ odd} \lsup{3}{{\cal K}}_\ell. \label{eq3:KHe4_3}
\end{align}
As total electronic spin is conserved, each long-range scattering
state converts to a single short-range molecular state associated
with a potential of Fig.~\ref{fig3:shortrangepot}. Consequently, the
strong suppression of ionization in the quintet potential results in
a rate coefficient $\lsup{5}{K} \approx 0$.

\subsubsection{Fermionic $^3\!$He*: Antisymmetric states with hyperfine structure}
In a collision of two $^3$He* atoms, the total atomic angular
momentum $\mathbf{F}=\mathbf{f}_1+\mathbf{f}_2$ and the relative
angular momentum of the two atoms $\bm{\ell}$ are, to good
approximation, conserved separately. Here,
$\mathbf{f}_j=\mathbf{s}_j+\mathbf{i}_j$ ($j=1,2$), the total
angular momentum $\mathbf{f}_j$ of an atom is the sum of its
electronic spin $\mathbf{s}_j$ and nuclear spin $\mathbf{i}_j$.

Ignoring the radial part of the quantum states and assuming for the
moment that the colliding atoms are distinguishable, a basis for
atom pairs is given by
\beq
\left\{ \rule[-2mm]{0mm}{4mm} |(s_1 i_1 f_1)_A (s_2 i_2 f_2)_B, F M_F, \ell m_\ell \rangle \right\},
\label{eq3:basislong}
\eeq
where atoms $A$ and $B$ have identical electronic spins,
$s_1=s_2=1$, and nuclear spins, $i_1=i_2=\frac{1}{2}$, that add up
to $f_1$ for atom $A$ and $f_2$ for atom $B$. The latter two add up
to the total atomic angular momentum $F$ with projection onto the
internuclear axis $M_F$. In magneto-optically trapped $^3$He*
samples, all atoms occupy the lower $f=\frac{3}{2}$ hyperfine level,
so that $f_1=f_2=\frac{3}{2}$ and $F$ can take values 0, 1, 2 and~3. Quantum numbers $\ell$
and $m_\ell$ are the angular momenta associated with the relative
motion of the two atoms and its projection onto the internuclear
axis, respectively.

As the $^3$He atom is a fermion, the physical states describing
atom pairs are antisymmetric under exchange of the atoms.
Applying antisymmetrizer \mbox{$\text{A}=(1 - \text{P}_{12})/\sqrt{2}$} to
the basis states \cite{cohe77,burk99t}, we obtain
\begin{multline}
|s_1 i_1 f_1,s_2 i_2 f_2, F M_F, \ell m_\ell \rangle  \\ =
 \frac{1}{\sqrt{2}} \left[1 + (-1)^{\ell-F}\right] \: |(s_1 i_1 f_1)_A (s_2 i_2 f_2)_B, F M_F, \ell m_\ell \rangle.
\end{multline}
These states, that must be normalized if necessary, differ from
zero only if $F-\ell$ is even. Consequently, only even (odd) partial
waves contribute to collisions with even (odd) $F$, and the total
ionization rate coefficient can be written
\beq
K(F) =  \begin{cases} \: \displaystyle\sum_\text{$\ell$ even} {\cal K}_\ell(F)& \text{if $F=0,2$}, \\
\rule[0mm]{0mm}{6mm} \: \displaystyle\sum_\text{$\ell$ odd} {\cal K}_\ell(F)& \text{if $F=1,3$}.
\end{cases} \label{eq3:KHe3}
\eeq
To express partial wave rate coefficient ${\cal K}_\ell(F)$ in terms
of the rate coefficients $\lsup{(2S+1)}{\cal K}_\ell$ that are
associated with singlet and triplet potentials, we consider the
interaction between two colliding $^3$He* atoms.

The collision process of two laser-cooled $^3$He* atoms is
controlled by the atomic hyperfine interaction \cite{rosn70} and the
various interatomic interactions \cite{mull91,yan98,spel93} that result in
potentials $\lsup{(2S+1)}{V}(R)$. As a result of the hyperfine interaction,
$S$ is not a good quantum number for large internuclear distances,
where atom pairs are characterized by $F$. However, $S$ is a good
quantum number for $R \lsim 30 \, a_0$, where the molecular
interaction dominates and Wigner's spin-conservation rule applies.
The evolution of the quantum mechanical state from long to short
internuclear distance is well approximated by a diabatic transition,
as the absolute change in the coupling between quasi-molecular
states, during one period of oscillation in the quasi-molecular
system, is much larger than the absolute energy difference between
the scattering states \cite{davy65,stas05t}. Around $R= 35 \, a_0$, the
molecular interaction, increasing exponentially with decreasing $R$
\cite{leo01}, becomes larger than the atomic hyperfine interaction,
while the relative velocity has increased to $4 \times
10^{-5}$~a.u.\ due to the attractive Van der Waals potential.
Consequently, the asymptotic scattering state of a colliding atom
pair remains unchanged when the atoms reach small internuclear
distance and the relation between rate coefficient ${\cal
K}_\ell(F)$ and coefficients $\lsup{(2S+1)}{\cal K}_\ell$ can be
determined by expanding the corresponding scattering state onto the
molecular states.

\begin{table*}
\caption{\label{table:pairstates}Expansion coefficients $a_{SI}(F) =
\langle s_1 i_1, s_2 i_2,S I, F M_F, \ell m_\ell | s_1 i_1 f_1,s_2
i_2 f_2, F M_F, \ell m_\ell \rangle$. Scattering state $|s_1 i_1
f_1,s_2 i_2 f_2, F M_F, \ell m_\ell \rangle$ is indicated by its
values of $F$, while molecular state $|s_1 i_1, s_2 i_2,S I, F M_F,
\ell m_\ell \rangle$ is denoted in the Hund's case (a) notation,
$^{2S+1}\Sigma_\text{g/u}^+$ ($I$).}
\begin{ruledtabular}
\begin{tabular}{lcccccc}
                & $^1\Sigma_\text{g}^+$ ($I=0$)     & $^1\Sigma_\text{g}^+$ ($I=1$)    & $^3\Sigma_\text{u}^+$ ($I=0$)    & $^3\Sigma_\text{u}^+$ ($I=1$)    & $^5\Sigma_\text{g}^+$ ($I=0$)    & $^5\Sigma_\text{g}^+$ ($I=1$) \rule[-2mm]{0mm}{6mm} \\ \hline
$F=0$           & $\sqrt{\frac{2}{3}}$              &                                  &                                  & $-\sqrt{\frac{1}{3}}$            &                                  &                         \rule[0mm]{0mm}{5mm} \\
$F=1$           &                                   & $\sqrt{\frac{10}{27}}$           & $\sqrt{\frac{5}{9}}$             &                                  &                                  & $-\sqrt{\frac{2}{27}}$  \rule[0mm]{0mm}{4mm} \\
$F=2$           &                                   &                                  &                                  & $\sqrt{\frac{2}{3}}$             & $\sqrt{\frac{1}{3}}$             &                         \rule[0mm]{0mm}{4mm} \\
$F=3$           &                                   &                                  &                                  &                                  &                                  &  1                      \rule[0mm]{0mm}{4mm} \\
\end{tabular}
\end{ruledtabular}
\end{table*}
The expansion coefficients only depend on the angular part of the
quantum states involved. For the molecular states, the angular part
can be derived applying the antisymmetrizer to basis
\beq
\left\{ \rule[-2mm]{0mm}{4mm} |(s_1 i_1)_A (s_2 i_2)_B, S I, F M_F, \ell m_\ell \rangle \right\}
\label{eq3:basisshort}
\eeq
(and normalizing if necessary), where the atoms $A$ and $B$ are
assumed distinguishable, $I$ is the quantum number associated with
the sum of the nuclear spins,
$\mathbf{I}=\mathbf{i}_1+\mathbf{i}_2$, and $S$ is the quantum
number associated with the sum of the electronic spins,
$\mathbf{S}=\mathbf{s}_1+\mathbf{s}_2$; the bases of
Eqs.~(\ref{eq3:basislong}) and (\ref{eq3:basisshort}) are related
through 9-$j$ symbols \cite{zare88}. The resulting physical states
\begin{multline}
|s_1 i_1, s_2 i_2,S I, F M_F, \ell m_\ell \rangle = \frac{1}{\sqrt{2}}[1+(-1)^{\ell-S-I}] \\
\times |(s_1 i_1)_A (s_2 i_2)_B, S I, F M_F, \ell m_\ell \rangle
\label{eq3:physstateshort}
\end{multline}
are different from zero only if $\ell-S-I$ is even. The expansion of
a scattering state onto the molecular states can be confined within
the subspace defined by $F$ and $\ell$, as $F$ and $\ell$ can be
considered good quantum numbers,
\begin{multline}
|s_1 i_1 f_1,s_2 i_2 f_2, F M_F, \ell m_\ell \rangle \\= \sum_{S,I} a_{SI}(F) \: |s_1 i_1, s_2 i_2,S I, F M_F,
\ell m_\ell \rangle.
\end{multline}
Table~\ref{table:pairstates} presents the expansion coefficients
$a_{SI}(F)$ for scattering states with $f_1=f_2=\frac{3}{2}$.

The partial wave rate coefficient ${\cal K}_\ell(F)$ associated with
scattering state $|s_1 i_1 f_1,s_1 i_1 f_2, F M_F, \ell m_\ell
\rangle$ can be written as a weighted sum of coefficients
$\lsup{(2S+1)}{\cal K}_\ell$ associated with molecular states $|s_1
i_1, s_1 i_1,S I, F M_F, \ell m_\ell \rangle$ with weights
$|a_{SI}(F)|^2$,
\beq
{\cal K}_\ell(F) = \sum_{S,I} |a_{SI}(F)|^2 \times \lsup{(2S+1)}{\cal K}_\ell.
\eeq
It can be seen in Table~\ref{table:pairstates} that, in the case of
$F=0$, only singlet and triplet states are involved, so that
\beq
{\cal K}_\ell(0)= \tfrac{2}{3} (\lsup{1}{\cal K}_\ell)
+ \tfrac{1}{3} (\lsup{3}{\cal K}_\ell). \label{eq3:exp0}
\eeq
For the $F=1$ and $F=2$ states, the diabatic transition transforms
the scattering states into a superposition of ionizing and
non-ionizing molecular states. The contribution to the ionization
rate coefficient from quintet states can be neglected, so that
\begin{align}
{\cal K}_\ell(1) &= \tfrac{10}{27} (\lsup{1}{\cal K}_\ell)
+ \tfrac{5}{9} (\lsup{3}{\cal K}_\ell), \\
{\cal K}_\ell(2) &= \tfrac{2}{3} (\lsup{1}{\cal K}_\ell).
\end{align}
Finally, in the case of $F=3$, partial rate coefficient ${\cal
K}_\ell(3)$ is negligible, as only quintet states are involved.

\subsection{\label{sec:th:unpol}Ionization rate coefficient for trapped samples}
\begin{table*}
\caption{\label{table:weights} Coefficients
$\lsup{(2S+1)}{b\onemu}_\text{even/odd}$ from Eq.~(\ref{eq:Karb}). The
coefficients are the sums of the expectation values of the density
operator for all ionizing molecular states of given $S$ and given parity.}
\begin{ruledtabular}
\begin{tabular}{lcc}
                            & $^3$He*         & $^4$He* \rule[-2mm]{0mm}{6mm} \\ \hline
$\lsup{1}{b\onemu}_\text{even}$   & $\frac{1}{3}(P_{-3/2}P_{3/2}+P_{-1/2}P_{1/2})$                & $\frac{1}{3}(2 P_{-1}P_1 + P_0^2)$                    \rule[0mm]{0mm}{5mm} \\
$\lsup{3}{b\onemu}_\text{even}$   & $\frac{2}{3}(P_{-3/2}P_{-1/2}+P_{-3/2}P_{1/2}+P_{-1/2}P_{3/2}+P_{1/2}P_{3/2}) + \frac{1}{2}(P_{-3/2}P_{3/2}+P_{-1/2}P_{1/2})$                & 0  \rule[0mm]{0mm}{4mm} \\
$\lsup{1}{b\onemu}_\text{odd}$    & $\frac{2}{9}(P_{-3/2}P_{1/2}+P_{-1/2}P_{3/2})+\frac{4}{27}(P_{-1/2}P_{-1/2}+P_{1/2}P_{1/2})+\frac{1}{3}P_{-3/2}P_{3/2}+\frac{1}{27}P_{-1/2}P_{1/2}$                & 0  \rule[0mm]{0mm}{4mm} \\
$\lsup{3}{b\onemu}_\text{odd}$    & $\frac{1}{3}(P_{-3/2}P_{1/2}+P_{-1/2}P_{3/2})+\frac{2}{9}(P_{-1/2}P_{-1/2}+P_{1/2}P_{1/2})+\frac{1}{2}P_{-3/2}P_{3/2}+\frac{1}{18}P_{-1/2}P_{1/2}$                & $P_{-1}P_0 + P_{-1}P_1 + P_0P_1$  \rule[0mm]{0mm}{4mm} \\
\end{tabular}
\end{ruledtabular}
\end{table*}
In a laser-cooled sample of He* atoms, collisions occur for all
values of the total atomic angular momentum, $F=0,1,2,3$ in case of
$^3$He*, and $S=0,1,2$ in case of $^4$He*. The contribution of each
collision channel depends on the distribution $P_m$ of magnetic
substates in the sample, where $m$ is the azimuthal quantum number
of the atom, which can take on values
$m_f=-\frac{3}{2},-\frac{1}{2}, \frac{1}{2},\frac{3}{2}$ in case of
$^3$He* and $m_s=-1,0,1$ in case of $^4$He*. Using density operator
\cite{cohe77}
\beq
\rho(\bm{r}) = \sum_m \sum_{n\leq m} P_m(\bm{r}) \, P_n(\bm{r}) \, |m, n \rangle \langle m,n |
\eeq
to describe a statistical mixture of (properly symmetrized) magnetic
substate pairs $|m,n\rangle$, with $m$ and $n$ azimuthal quantum
numbers, the ionization rate coefficient for the mixture (cf.\ Eq.~(\ref{eq:ionrate})) can be
written as
\begin{multline}
K = \frac{1}{N} \iiint \Bigl[ \sum_{\text{$\ell$ even}} \bigl( \lsup{1}{b\onemu}_\text{even} \: \lsup{1}{\cal K}_\ell
+ \lsup{3}{b\onemu}_\text{even} \: \lsup{3}{\cal K}_\ell \bigr)\\
+ \sum_{\text{$\ell$ odd}} \bigl( \lsup{1}{b\onemu}_\text{odd} \: \lsup{1}{\cal K}_\ell
+ \lsup{3}{b\onemu}_\text{odd} \: \lsup{3}{\cal K}_\ell \bigr)\Bigr] n(\bm{r}) \, \text{d}^3r. \label{eq:Karb}
\end{multline}
where $N$ is the number of trapped atoms, $n(\bm{r})$ is the density
distribution in the sample, and
$\lsup{(2S+1)}{b\onemu}_\text{even/odd}$ is the sum of the
expectation values of the density operator for all ionizing
molecular states with total spin $S$ and even/odd parity. Explicit
expressions for the coefficients are given in
Table~\ref{table:weights}. As the coefficients can be interpreted as
projections of the statistical mixture onto subspaces of ionizing
states, Eq.~(\ref{eq:Karb}) implies the assumption that
scattering states transform diabatically to a superposition of
molecular states, when atoms move from large to small internuclear
distance.

\begin{figure}[t]
\begin{center}
\includegraphics[width=\columnwidth,keepaspectratio]{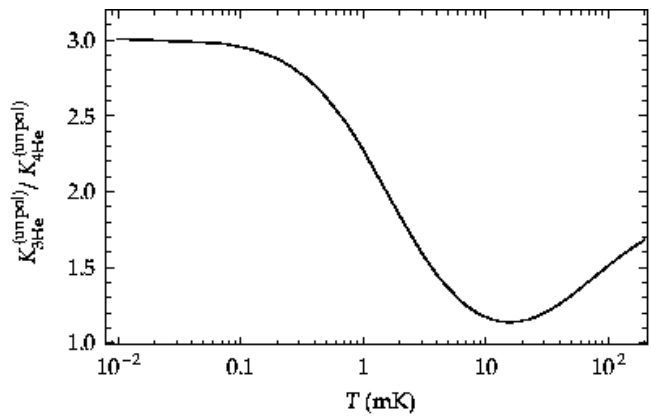}
\caption{\label{fig3:bunpolratio}Ratio of unpolarized ionization
rate coefficients
$K_\text{3He}^{\text{(unpol)}}/K_\text{4He}^{\text{(unpol)}}$,
including partial wave contributions up to $\ell=2$.}
\end{center}
\end{figure}
The unpolarized ionization rate coefficients
$K^{\text{(unpol)}}$, i.e.\ the rate coefficient for a
laser-cooled sample of He* atoms where the magnetic substates of the
atoms are uniformly populated, is obtained by setting
$P_{-1}=P_0=P_1=\frac{1}{3}$ for $^4$He* or
$P_{-3/2}=P_{-1/2}=P_{1/2}=P_{3/2}=\frac{1}{4}$ for $^3$He*. For
samples with a temperature around 1~mK, only s and p-waves have to
be taken into account. For unpolarized $^4$He* samples, we obtain
\beq
K_\text{4He}^{\text{(unpol)}} \approx \tfrac{1}{9} \left[ (\lsup{1}{{\cal K}}_0) + 3
(\lsup{3}{{\cal K}}_1) \right],
\eeq
where $\lsup{1}{\cal K}_0$ and $\lsup{3}{\cal K}_1$ are the partial
wave ionization rate coefficients for $^4$He*, calculated in
Sec.~\ref{sec:th:num}. In the case of an unpolarized sample of
$^3$He* atoms in the lower $f=\frac{3}{2}$ hyperfine level, the rate
coefficient is given by
\beq
K_\text{3He}^{\text{(unpol)}} \approx \tfrac{1}{16} \left[\tfrac{11}{3}(\lsup{1}{\cal K}_0) + \tfrac{2}{3}
(\lsup{3}{\cal K}_0) + \tfrac{10}{9} (\lsup{1}{\cal K}_1) + \tfrac{5}{3} (\lsup{3}{\cal K}_1) \right],
\eeq
where $\lsup{1}{\cal K}_0$, $\lsup{3}{\cal K}_0$, $\lsup{1}{\cal
K}_1$ and $\lsup{3}{\cal K}_1$ are the partial wave ionization rate
coefficients for $^3$He*. For both isotopes, the coefficients are
shown in Fig.~\ref{fig3:bunpol} for temperatures between
10~$\micro$K and 100~mK. It can be seen in
Fig.~\ref{fig3:bunpolratio}, where the ratio of the unpolarized rate
coefficients for the two isotopes is displayed, that unpolarized
$^3$He* atoms have a larger rate coefficient than unpolarized
$^4$He* atoms for temperatures between 10~$\micro$K and 100~mK.

\begin{table}
\caption{\label{tab3:theorycompare}Calculated values of $K^{\text{(unpol)}}$ and comparison between various theoretical results.
Theoretical results from Ref.~\cite{leo01} are extracted from Fig.~8 of that paper.}
\begin{ruledtabular}
\begin{tabular}{lcccc}
        &Ref.           & T (mK)        & $K^{\text{(unpol)}} (\text{cm$^3$/s})$ & This work             \\ \hline
$^4$He* &\cite{leo01}   & 0.001         & $9.9 \times 10^{-11}$         & $7.7 \times 10^{-11}$ \\
        &\cite{kuma99}  & 0.5           & $2.2 \times 10^{-10}$         & $7.5 \times 10^{-11}$ \\
        &\cite{leo01}   & 0.5           & $8.6 \times 10^{-11}$         & $7.5 \times 10^{-11}$ \\
        &\cite{mast98}  & 1             & $7.3 \times 10^{-11}$         & $8.3 \times 10^{-11}$ \\
        &\cite{leo01}   & 1             & $8.9 \times 10^{-11}$         & $8.3 \times 10^{-11}$ \\
        &               & 2             &                               & $9.9 \times 10^{-11}$ \\ \hline
$^3$He* &\cite{kuma99}  & 0.5           & $1.0 \times 10^{-9}$          & $2.0 \times 10^{-10}$ \\
        &               & 2             &                               & $1.8 \times 10^{-10}$ \\
\end{tabular}
\end{ruledtabular}
\end{table}
In Table~\ref{tab3:theorycompare}, we compare the results of the
theoretical model presented here to results of other theoretical
work \cite{mast98,kuma99,leo01}. We see that our results agree well
with the results of the detailed close-coupling theory of
\cite{leo01} and the simpler calculation of \cite{mast98}, but that
there is a large discrepancy with the results of Kumakura and Morita
\cite{kuma99}. This is not surprising, since their model does not
account for quantum reflection for s-wave scattering. As we have
shown in Sec.~\ref{sec:th:num}, quantum reflection is significant
and we estimate that the omission of this effect leads to rate
coefficients that are too large by factor of about~2. Moreover,
Kumakura and Morita assume that the evolution of the scattering
states during the collision of two $^3$He* atoms can be approximated
by an adiabatic transition and accordingly apply the noncrossing rule
\cite{niki84} to derive the number of ionization channels for each
partial wave ionization rate coefficient. As shown in
Sec.~\ref{sec:th:symm}, the system is well approximated by a
diabatic transition and we estimate that the assumptions of an
adiabatic transition leads to an unpolarized rate coefficient that
is 50\% too large. This explains the difference
between our results and those of Ref.~\cite{kuma99}.
As a final remark we note that the observed differences between our work and those of Refs. \cite{mast98} and \cite{leo01} may be
due to theoretical uncertainties in the molecular potentials and in the form of the ionization widths used in the calculations.
This leads to theoretical uncertainties of $\approx$40\%, as discussed in Ref. \cite{leo01}.

\section{\label{sec:exp}Measurement of ionizing collisions in a magneto-optical trap}
The experimental investigation of ionizing collisions is performed
in a setup that can be used to confine large numbers ($\gsim 10^8$)
of atoms in a \MOT. As the setup has been described previously
\cite{stas04}, only a brief outline is presented here. The \MOT\ is
loaded from a collimated and Zeeman slowed atomic beam, produced by
a liquid nitrogen cooled, \textsc{dc}-discharge source. The beam
source can be operated with pure $^3$He* or pure $^4$He* gas. In the
case of $^3$He*, the gas is recycled and purified using liquid
nitrogen cooled molecular sieves.

The laser light that is used for collimation, Zeeman slowing and
magneto-optical trapping has a wavelength of 1083~nm and is near
resonant with the $2~\lsup{3}{\text{S}}_1 (f=\frac{3}{2})
\rightarrow 2\: \lsup{3}{\text{P}}_2 (f=\frac{5}{2})$ optical
transition in the case of $^3$He*, and the $2~^3\text{S}_1
\rightarrow 2~^3\text{P}_2$ transition in the case of $^4$He*. For both
transitions, the natural linewidth $\Gamma/2 \pi = 1.62$~MHz and the
saturation intensity $I_{\text{sat}} = 0.167$~mW/cm$^2$ (cycling
transition). The light is generated by an ytterbium-doped fiber
laser, that is frequency-stabilized to the laser cooling transition
using saturated absorption spectroscopy in an rf-discharge cell.
Acousto-optic modulators (\AOM s) are used to generate the slowing
and trapping frequencies, which are detuned by $-500$~MHz and
$-40$~MHz, respectively. The slowing beam is focused onto the atomic
beam source, while the trapping light is split up into six
independent Gaussian beams with rms diameters of 27~mm and a total
peak intensity $I_{\text{peak}}=59$~mW/cm$^2$
($I_{\text{peak}}/I_{\text{sat}}\approx 353$). The magnetic
quadrupole field of the \MOT\ is generated by two anti-Helmholtz
coils and has an axial field gradient $\partial B/\partial z =
0.28$~T/m.

For the investigation of ionizing collisions, the trapped He*
samples are studied with an absorption imaging system and two
microchannel plate (\MCP) detectors \cite{stas05t}. The absorption
imaging system consists of a \CCD\ camera, a narrowband commercial diode laser
at 1083~nm (TOPTICA, model DL100) and an \AOM. The laser and \AOM\ are used to generate
low-intensity ($I \lsim 0.05 I_\text{sat}$) probe light pulses with
a duration of 100~$\micro$s, while the \CCD\ camera (mounted behind
a magnifier lens) records absorption images of the sample with a
magnification of~0.17. The images are used to determine the
(Gaussian) density distribution of the atoms in the trap
\cite{stas05t},
\beq
n(x,y,z,t) = n_0(t) \, \exp \! \left( - \frac{x^2}{2 \sigma_\rho^2} -
\frac{y^2}{2 \sigma_\rho^2} - \frac{z^2}{2 \sigma_z^2} \right),
\label{eq5:gaussian}
\eeq
with $n_0(t)$ the (time-dependent) density in the center ($x=y=z=0$)
of the sample, and $\sigma_\rho$ and $\sigma_z$ the rms
radii of the distribution; the number of trapped atoms $N=\iiint
n(\bm{r}) \, \text{d}^3 r = (2 \pi)^{3/2} \sigma_\rho^2 \sigma_z
\, n_0$.

The \MCP\ detectors allow for the independent monitoring of ions and
He* atoms that escape the trap. With an exposed negative high
voltage on its front plate, one \MCP\ detector attracts all ions
produced in the trapped sample. The other \MCP\ is mounted behind a
grounded grid and detects only the He* atoms that exit the trap in its
direction. This shielded \MCP\ detector is used to determine the
temperature $T$ of the trapped samples through time-of-flight
measurements. The unshielded \MCP\ detector measures the instantaneous
ionization rate in the trapped sample and is used to determine trap
loss and ionization rates in the sample.

\subsection{\label{sec:exp:loss}Trap loss and ionization in magneto-optical trap}
For magneto-optically trapped He* samples, the time-evolution of the
number of trapped atoms $N$ can be described by the phenomenological
equation \cite{bard92,brow00}
\beq
\frac{\text{d}N(t)}{\text{d}t}=L - \alpha N(t) - \beta \iiint \! n^2(\bm{r},t) \, \text{d}^3 r.
\label{eq5:diffnumb}
\eeq
Writing down this differential equation, we assume that $N(t)$ is
controlled by three simultaneously occurring processes,
corresponding to the three terms on the right-hand side of the
equation. The first term is a constant loading rate $L$,
representing the capture of atoms from the decelerated atomic beam
into the \MOT. The second and third term are the linear and
quadratic trap loss rate, respectively. The loss processes are
defined in terms of the local atomic density of the sample
$n(\bm{r},t)$ (with $L=0$) by
\beq
\frac{\text{d}n(\bm{r},t)}{\text{d}t}= - \alpha \, n(\bm{r},t) - \beta \, n^2(\bm{r},t).
\label{eq5:diffdens}
\eeq
The nomenclature of the loss rate terms refers to their
proportionality to density $n(\bm{r},t)$ and density squared
$n^2(\bm{r},t)$. Analogously, the proportionality constants,
$\alpha$ and $\beta$, are referred to as the linear and quadratic
loss rate coefficient, respectively.

For He* samples in a 1083~nm \MOT, only collisional loss mechanisms
give rise to significant trap loss. Quadratic trap loss is
determined by collisions between trapped He* atoms, while linear
trap loss results from collisions with particles traversing the
trapping volume, such as background gas particles and helium atoms
from the atomic beam. Loss rates in $^4$He* samples can be estimated
using cross section data reported in literature
\cite{yenc84,sisk93}. Table~\ref{table:lossrates} presents an
overview of trap loss mechanisms in $^4$He* samples; cross section
data are combined with the atomic density in the sample, the
background gas density or the atomic beam flux to determine the
estimates.

The trap loss mechanisms can be subdivided into ionizing
mechanisms and mechanisms where atoms are lost without the formation
of ions. Linear trap loss is dominated by a non-ionizing loss
mechanism. As the beam of metastable atoms is not separated from the
beam of ground state atoms (contrary to other work
\cite{tol99,hers00,brow00,pere01a}), collisions of ground state
$^4$He atoms from the atomic beam with trapped atoms (collision
energy $E_\text{r} \approx 4.9$~meV) give rise to a trap loss rate
of about $2$~s$^{-1}$. Ionizing mechanisms hardly contribute:
collisions with slowed ($E_\text{r} \approx 0.0064$~meV) and
non-slowed $^4$He* atoms ($E_\text{r} \approx 6.5$~meV) from the
atomic beam, and collisions with background molecules (presumably
H$_2$O, H$_2$, N$_2$ and O$_2$) result in a loss rate of about $1
\times 10^{-2}$~s$^{-1}$.

Quadratic trap loss is dominated by ionization mechanisms. In the
presence of trapping light, light-assisted ionizing collisions
between trapped $^4$He* atoms give rise to an ionization rate of
4~s$^{-1}$. In the absence of trapping light, ionizing collisions
yield a rate of 0.1~s$^{-1}$. As two atoms are lost for every ion
that is formed, the corresponding trap loss rates are 8~s$^{-1}$ and
0.2~s$^{-1}$, respectively. The escape of fast $^4$He* atoms from
the trap, through fine-structure-changing collisions or radiative
escape, constitutes a non-ionizing quadratic loss mechanism in the
presence of trapping light, that can be neglected
\cite{hers03t,koel04}.

The ionization rate associated with the linear and quadratic
ionization mechanisms can be written
\beq
\frac{\text{d}N_\text{ion}(t)}{\text{d} t} = \epsilon_a \alpha N(t)
+ \frac{\epsilon_b \beta}{2} \iiint \! n^2(\bm{r},t) \, \text{d}^3r,
\label{eq5:ioniz}
\eeq
where $\epsilon_a$ and $\epsilon_b$ are the weights of ionization
mechanisms in linear and quadratic trap loss, respectively. The
quadratic ionization rate is half of the ionizing quadratic trap
loss rate, as a single ion is formed for every pair of colliding He*
atoms that is lost from the trap. As the linear ionization rate is
small compared to the quadratic ionization rate, both in the
presence and absence of trapping light, and quadratic trap loss is
almost completely determined by ionization mechanisms, we can set,
to good approximation, $\epsilon_a = 0$ and $\epsilon_b = 1$, so
that
\beq
\frac{\text{d}N_\text{ion}(t)}{\text{d} t} \approx \frac{\beta}{2} \iiint \! n^2(\bm{r},t) \, \text{d}^3r.
\label{eq5:ionizapprox}
\eeq

For samples of $^3$He* atoms, trap loss and ionization are
determined by the same loss mechanisms. As collision studies are
rare for $^3$He* atoms \cite{yenc84,sisk93}, we cannot derive
accurate estimates of trap loss and ionization rates in this case.
However, cross sections are not expected to show a large isotopic
dependence (less than a factor of two), so we conclude that
Eq.~(\ref{eq5:ionizapprox}) also applies to $^3$He* samples. It
should be noted in this respect, that hyperfine-changing collisions
are forbidden by energy conservation. Trapped $^3$He* atoms are in
the lower $f=\frac{3}{2}$ hyperfine level, so that
hyperfine-changing collisions require an energy larger than the
hyperfine splitting $E_\text{hf}=28~\micro$eV, which corresponds to
a temperature $E_\text{hf}/ \frac{3}{2} k_\text{B}=0.2$~K.
\begin{table*}
\caption{\label{table:lossrates} Trap loss rate
$\text{d}N/\text{d}t$ and ionization rate
$\text{d}N_\text{ion}/\text{d}t$ associated with collisional loss
mechanisms in magneto-optically trapped samples of $^4$He* atoms.
Rates are calculated for collisions of trapped $^4$He* atoms with
ground state ($1\; \lsup{1}{\text{S}}$) $^4$He atoms and metastable
($2\; \lsup{3}{\text{S}}$) $^4$He* atoms from the atomic beam, for
collisions with ground state $^4$He atoms and several molecules from
the thermal background gas, and for collisions between trapped
$^4$He* atoms from the sample, both in the presence and absence of
trapping light (with a wavelength of 1083~nm). For each collisional
mechanism, collision energy $E_r$ is given, as well as total cross
section $\sigma^\text{(tot)}$ or ionization cross section
$\sigma^\text{(ion)}$. Furthermore, beam flux $F$, background gas
density $\tilde{n}$, or central density $n_0$ of the sample are
given. For collisions with $^4$He* atoms from the beam, we
distinguish between slowed ($E_r= 0.0064$~meV) and non-slowed ($E_r=
6.5$~meV) atoms \cite{stas05t}.
The cross sections are taken from various references; in references
where an ionization rate coefficient $K$ is given the corresponding
cross section is calculated as $\sigma^\text{(ion)} =
K/\bar{v}_\text{r}$, where $\bar{v}_\text{r}=(8 k_\text{B} T/\pi
\mu)^{1/2}$ is the mean relative velocity for collisions of
particles with mass $m_1$ and $m_2$ in a gas with temperature $T$,
with reduced mass $\mu = m_1 m_2 / (m_1+m_2)$.}
\begin{ruledtabular}
\begin{tabular}{llccccccccc}
Collision                        & Originating & $E_\text{r}$    & $\sigma^\text{(tot)}$   & $\sigma^\text{(ion)}$                   & $F$                                       & $\tilde{n}$           & $n_0$             & $\text{d}N_\text{ion}/\text{d}t$    & $\text{d}N/\text{d}t$\\
\, partner                       & \multicolumn{1}{c}{from\:}        & (meV)           & ($10^{-16}\onemu$cm$^2$)     & ($10^{-16}\onemu$cm$^2$)                     & (cm$^{-2}\,$s$^{-1}$)                      & (cm$^{-3}$)           & (cm$^{-3}$)       & (s$^{-1}$)                          & (s$^{-1}$) \\ \hline
$^4$He                           & beam        & 4.9             & 200\onemu\footnotemark[1] & ---                                  & $\sim 10^{14}$                            & ---               & ---               & 0                                       & $\sim 2$\onemu\footnotemark[2] \rule[0mm]{0mm}{4mm} \\
$^4$He*                          & beam        & 6.5             & ---                     & 181\onemu\footnotemark[3]               & $4 \times 10^{11}$                   & ---                   & ---               & $7 \times 10^{-3}$\,\footnotemark[2]   & $1 \times 10^{-2}$\,\footnotemark[4] \\
$^4$He*                          & beam        & 0.0064          & ---                     & $1160$\onemu\footnotemark[5]           & $8 \times 10^{9}$                     & ---                   & ---               & $9 \times 10^{-4}$\,\footnotemark[2]   & $2 \times 10^{-3}$\,\footnotemark[4] \\
$^4$He                           & background  & 16.5            & 140\onemu\footnotemark[6] & ---                                  & ---                                   & $1 \times 10^7$       & ---               & 0                                       & 0.02\onemu\footnotemark[7] \\
H$_2$                            & background  & 22              & ---                     & 0.1\onemu\footnotemark[8]               & ---                                       & $< 1.2 \times 10^6$   & ---           & $<2 \times 10^{-4}$\,\footnotemark[7]  & $<4 \times 10^{-4}$\,\footnotemark[4] \\
H$_2$O                           & background  & 6.0             & ---                     & 131\onemu\footnotemark[9]               & ---                                       & $< 1.2 \times 10^6$   & ---           & $<9 \times 10^{-4}$\,\footnotemark[7]  & $<2 \times 10^{-3}$\,\footnotemark[4] \\
O$_2$                            & background  & 4.7             & ---                     & 8\onemu\footnotemark[10]                 & ---                                       & $< 1.2 \times 10^6$   & ---           & $<4 \times 10^{-5}$\,\footnotemark[7]  & $<8 \times 10^{-5}$\,\footnotemark[4] \\
N$_2$                            & background  & 4.1             & ---                     & 2\onemu\footnotemark[11]                 & ---                                       & $< 1.2 \times 10^6$   & ---           & $<1 \times 10^{-5}$\,\footnotemark[7]  & $<2 \times 10^{-5}$\,\footnotemark[4] \\
$^4$He*                          & sample      & 0.00001         & ---                     & $4 \times 10^3$\,\footnotemark[12]  & ---                                       & ---                   & $4 \times 10^9$   & 0.1\onemu\footnotemark[13]              & 0.2\onemu\footnotemark[4] \\
$^4\text{He*} + \, \text{light}$ & sample      & 0.00001         & ---                     & $16 \times 10^4$\,\footnotemark[12] & ---                                       & ---                   & $4 \times 10^9$   & 4\onemu\footnotemark[13]                & 8\onemu\footnotemark[4] \\
\end{tabular}
\end{ruledtabular}
\footnotetext[1]{Vrinceanu \emph{et al.\ }\cite{vrin02}.}
\footnotetext[2]{For collisions with atoms from the beam, rates are $\sigma F$.}
\footnotetext[3]{For collision energies $0.1\text{~meV}< E_\text{r} < 100\text{~meV}$,
$\sigma^\text{(ion)}$ satisfies the (semi-classical) energy
dependence $\sigma^\text{(ion)} = \sigma_1 (E_1/E_\text{r})^\alpha$, with
$E_1=1$~meV, $\sigma_1=318 \times 10^{-16}$~cm$^2$ and $\alpha=0.3$
\cite{mull91}.}
\footnotetext[4]{As two atoms are lost from the trap for each ion that is formed, $\text{d}N/\text{d}t=2\bigl(\text{d}N_\text{ion}/\text{d}t\bigr)$.}
\footnotetext[5]{Venturi \emph{et al.\ }\cite{vent00}.}
\footnotetext[6]{Mastwijk \cite{mast97t} and Rothe \emph{et al.\ }\cite{roth65}.}
\footnotetext[7]{For collisions with background gas particles,
rates are $\sigma \bar{v}_\text{r} \tilde{n}$, where $\bar{v}_\text{r}= (2 E_\text{r}
/ \mu)^{1/2}$ is the mean relative velocity for collisions of
particles with mass $m_1$ and $m_2$, with $\mu = m_1 m_2 / (m_1+m_2)$
their reduced mass.}
\footnotetext[8]{Martin \emph{et al.\ }\cite{mart89}.}
\footnotetext[9]{Mastwijk \cite{mast97t}; the large cross section results from a
large attractive force between $^4$He* and H$_2$O, a consequence of
the permanent dipole moment of the H$_2$O molecule.}
\footnotetext[10]{Parr \emph{et al.\ }\cite{parr82}.}
\footnotetext[11]{Yamazaki \emph{et al.\ }\cite{yama02}.}
\footnotetext[12]{Tol \emph{et al.\ }\cite{tol99}; the cross section for light-assisted collisions is exceptionally
large due to the optical excitation of the colliding atom pair to
long-range dipole-dipole potentials \cite{wein03}.}
\footnotetext[13]{For samples with a Gaussian density distribution, as given by
Eq.~(\ref{eq5:gaussian}), quadratic ionization rate
$\text{d}N_\text{ion}/\text{d}t = \bigl(\sigma^\text{(ion)}
\bar{v}_\text{r}/N \bigr) \iiint n^2(\bm{r}) \, \text{d}^3r=
\sigma^\text{(ion)} \bar{v}_\text{r} n_0 / 2 \sqrt{2}$.}
\end{table*}

\subsection{\label{sec:exp:light}Ionization rates for light-assisted collisions}
To determine the ionization rate coefficient for light-assisted
collisions, $\epsilon_b \beta/2 \approx \beta/2$, we have performed
a trap loss experiment, where the loading of atoms to the \MOT\ is
abruptly stopped and the decaying ionization rate in the sample is
monitored with the unshielded \MCP\ detector. For Gaussian samples,
described by Eq.~(\ref{eq5:gaussian}), the ionization rate can be
written as
\beq
\frac{\text{d}N_\text{ion}(t)}{\text{d} t} = V \frac{\beta}{4 \sqrt{2}} \, n_0^2(t),
\eeq
where effective volume $V=(2 \pi)^{3/2} \sigma_\rho^2 \sigma_z$,
such that central density $n_0(t)=N(t)/V$. The current signal from
the \MCP\ detector is proportional to the ionization rate and is
converted to a voltage signal that is given by
\beq
\varphi(t) = e R_\text{eff} \frac{\beta}{4 \sqrt{2}} \, n_0^2(t), \label{eq5:ioncurr}
\eeq
where $e$ is the electron charge and $R_\text{eff}$ is an effective
resistance.

Substitution of Eq.~(\ref{eq5:gaussian}) into
Eq.~(\ref{eq5:diffnumb}) shows that the central density satisfies
differential equation
\beq
\frac{\text{d}n_0(t)}{\text{d}t} = \frac{L}{V}- \alpha n_0(t) -
\frac{\beta}{2 \sqrt{2}} \, n_0^2(t).
\label{eq5:diffn0}
\eeq
If the loading of atoms to the \MOT\ is abruptly stopped, the
density decays as \cite{pren88}
\beq
n_0(t) = \frac{n_0(0)}{\left( 1+\dfrac{\beta n_0(0)}{2 \sqrt{2}
\alpha} \right) \exp(\alpha t) - \dfrac{\beta n_0(0)}{2 \sqrt{2}
\alpha}}.
\label{eq5:evoln0}
\eeq
Substitution of Eq.~(\ref{eq5:evoln0}) into Eq.~(\ref{eq5:ioncurr})
gives an expression for the decaying ionization signal. Loss rates
$\alpha$ and $\beta n_0(0)$ determine the exact shape of the decay
and are derived from the measured traces by means of curve fitting.

\begin{figure}[t]
\begin{center}
\includegraphics[width=\columnwidth,keepaspectratio]{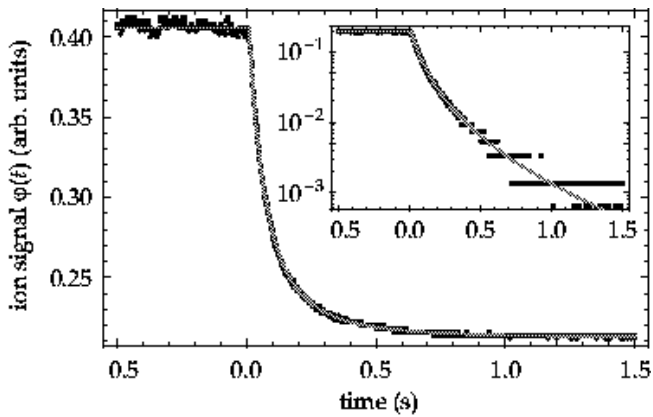}
\caption{\label{fig5:decayhe3}Ionization signal for a trap loss
measurement on a $^3$He* sample. At $t=0$, the input of atoms to the
\MOT\ is stopped abruptly. The rapid decay of the ionization signal
(black dots) is non-exponential, as can be seen in the inset. Fitting the
signal to our model (grey line) yields trap loss rates $\beta n_0(0)
= 20$~s$^{-1}$ and $\alpha = 0.7$~s$^{-1}$.}
\end{center}
\end{figure}
\begin{table}
\caption{\label{tab5:expres}Characteristic parameters of the magneto-optically
trapped He* samples and extracted loss rate coefficients. Experimental errors correspond to one standard
deviation.}
\begin{ruledtabular}
\begin{tabular}{lcc}
                                        & $^3$He*                       & $^4$He* \\ \hline
$T$ (mK)                                & 2.0(3)                        &$1.9(1)$ \rule[0mm]{0mm}{4mm} \\
$N$ ($-$)                               & $2.6(9) \times 10^{8}$        & $3.7(5) \times 10^{8}$ \\
$n_0$ (cm$^{-3}$)                       & $3.0(5) \times 10^9$          & $4.4(4) \times 10^9$ \\
$\alpha$ (s$^{-1}$)                     & 0.8(2)                        & 0.6(3) \\
$\beta$ (cm$^{3}$/s)                    & $5.5(8) \times10^{-9}$        & $3.3(7) \times10^{-9}$ \\
$K$ (cm$^{3}$/s)                        & $1.8(3) \times10^{-10}$       & $8(2) \times10^{-11}$ \\
$K^{\text{(unpol)}}$ (cm$^{3}$/s)       & $1.9(3) \times10^{-10}$       & $10(2) \times10^{-11}$ \\

\end{tabular}
\end{ruledtabular}
\end{table}
We have performed trap loss measurements on $^3$He* and $^4$He*
samples. The loading of atoms into the \MOT\ is stopped by blocking
the Zeeman slowing light with the \AOM\ used for frequency detuning
the slowing light from the atomic transition. Decay traces are
averaged four times using a digital oscilloscope. It has been
verified that the variations in the central density are small enough
that an averaged decay curve allows an accurate determination of
loss rates $\alpha$ and $\beta n_0(0)$. An example of an averaged
decay trace and the corresponding fit are displayed in
Figure~\ref{fig5:decayhe3}. The central density in the samples
$n_0(0)$ is derived from absorption imaging. The resulting rate
coefficients $\alpha$ and $\beta$ are presented in
Table~\ref{tab5:expres}.

The linear loss rate coefficients are close to 1~s$^{-1}$ for both
isotopes. Assuming that the loss rate stems from collisions with
ground state atoms from the atomic beam, we can determine the
intensity of the beam of ground state atoms. Using the total cross
section $\sigma = 200 \times 10^{-16}$~cm$^2$ \cite{vrin02} for
collisions between trapped $^4$He* atoms and ground state $^4$He
atoms from the atomic beam, the distance between source and trapped
sample of 370~cm, and loss rate coefficient $\alpha =
0.6$~s$^{-1}$, we derive an intensity of $4 \times
10^{18}$~s$^{-1}$~sr$^{-1}$. As the intensity of $^4$He* atoms is $4
\times 10^{14}$~s$^{-1}$~sr$^{-1}$ \cite{stas05t}, the fraction of
$^4$He* atoms in the beam is $10^{-4}$.

The quadratic loss rate coefficient for $^3$He* is almost twice as
large as the loss rate coefficient for $^4$He*. It has been pointed
out that this isotopic difference stems from a difference in the
relative number of ionization channels, opened up by the lowering of
centrifugal barriers by the long-range, resonant dipole-dipole
interaction \cite{kuma99}. Quantum statistical symmetry requirements
play a role in these collisions, but the effects are washed out as
the number of participating partial waves is much larger than one.
The loss rate coefficient for $^4$He* is in good agreement with
other work \cite{tol99}.

\subsection{\label{sec:exp:dark}Ionization in the absence of trapping light}
Collisions where ionization is not preceded by absorption of
trapping light contribute little to the ionization rate in
magneto-optically trapped He* samples. However, the corresponding
ionization rate coefficients can be determined from a comparative
measurement of the ionization rate in the presence and absence of
trapping light \cite{bard92}. In this measurement, the trapping and
slowing light are blocked for a short time interval of 100~$\micro$s
with the \AOM s used for the detuning of the laser frequencies from
the atomic transition. With the trapping light present, the observed
ionization signal
\beq
\varphi_\text{on}(t) = e R_\text{eff} \frac{\beta}{4 \sqrt{2}} \, n_0^2(0) + \varphi_\text{bgr}
\label{eq:phi_light}
\eeq
is dominated by light-assisted collisions and is relatively large. If the
trapping light is absent, collisions of $^3$He* or $^4$He* atoms
occur without optical excitation; the corresponding ionization rate
is much smaller and can be written as
\beq
\varphi_\text{off}(t) = e R_\text{eff} \frac{K}{2 \sqrt{2}} \, n_0^2(0) + \varphi_\text{bgr},
\label{eq:phi_dark}
\eeq
with $K$ the ionization rate coefficient in the absence of
trapping light. Combing Eqs.~(\ref{eq:phi_light}) and
(\ref{eq:phi_dark}), the rate coefficient can be written as
\beq
K = \frac{\beta}{2} \, \frac{\varphi_\text{off}-\varphi_\text{bgr}}{\varphi_\text{on}-\varphi_\text{bgr}}.
\eeq
Clearly, $K$ can be derived from a measurement of the ionization
rate coefficient for light-assisted collisions, $\beta/2$, and
signals $\varphi_\text{on}$, $\varphi_\text{off}$ and
$\varphi_\text{bgr}$.

\begin{figure}[t]
\begin{center}
\includegraphics[width=\columnwidth,keepaspectratio]{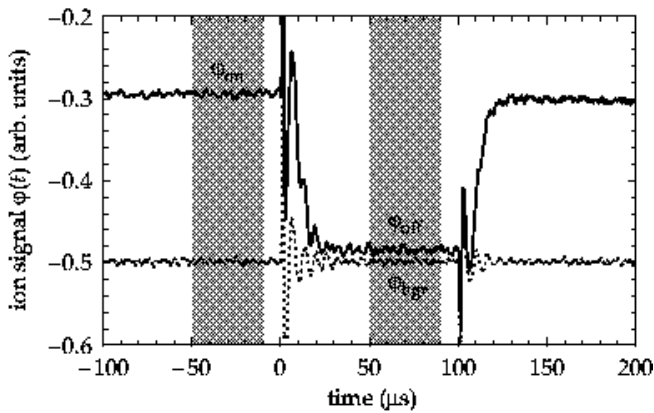}
\caption{\label{fig5:ksshe3}Ionization rates in the presence and
absence of trapping (and slowing) light. The signals are averaged
over 40~$\micro$s time intervals indicated by the shaded areas to
obtain $(\varphi_\text{on}-\varphi_\text{bgr})$ and
$(\varphi_\text{off}-\varphi_\text{bgr})$.}
\end{center}
\end{figure}
In the experiment, the trapping and slowing light are blocked every
200~ms, and the ionization signal is averaged 256 times with a
digital oscilloscope. Examples of averaged ionization signals are
displayed in Figure~\ref{fig5:ksshe3}. The measurement is repeated
with the atomic beam blocked to obtain the background ionization
signal $\varphi_\text{bgr}$, that includes an offset originating
from the \MCP\ signal amplifier. To obtain accurate values of
$(\varphi_\text{on}-\varphi_\text{bgr})$ and
$(\varphi_\text{off}-\varphi_\text{bgr})$, the average of the
signals is determined over 40~$\micro$s time intervals, as indicated
in Figure~\ref{fig5:ksshe3}.

Although the atoms are not confined if the trapping light is absent,
the expansion of the trapped He* sample during 100~$\micro$s is
insignificant and can be neglected. It is checked that the trapped
sample remains unaffected if the light is blocked repeatedly,
running the experiment with a duty cycle of 200~ms. The resulting
ionization rate coefficients are presented in
Table~\ref{tab5:expres}. The coefficients are close to but
do not agree within the error bars with other
experimental results \cite{mast98,kuma99,tol99,hers00}, that suffer
from mutual inconsistency themselves, as has been pointed out
\cite{leo01}.

\subsection{\label{sec:exp:comp}Comparison between experiment and theory}
To confront the theoretical model of Sec.~\ref{sec:th} with the
measurements of Secs.~\ref{sec:exp:light} and \ref{sec:exp:dark}, we
compare the theoretical expression for $K$, Eq.~(\ref{eq:Karb}),
with the $T=2$~mK experimental values of Table~\ref{tab5:expres}. As optical
pumping processes cause the magnetic substate distribution to
deviate from a uniform distribution (with the stretched substates
slightly overpopulated), we need to determine the magnetic substate
distribution $P_m(\bm{r})$ in the trapped samples to calculate a
value for $K$ that can be compared to experiment.

Distribution $P_m(\bm{r})$ is obtained as the steady-state solution
of a rate equation model, that describes the optical pumping in the
\MOT\ \cite{hers00}. Starting from the intensity and detuning of the trapping
light, and an expression for the quadrupole magnetic field, rate
equations are formulated and subsequently solved to obtain the
steady state substate population in the sample. In these
calculations we take into account that the intensities of the
trapping beams are not balanced and that (consequently) the trapped
sample is not exactly centered on the zero point of the magnetic
field.

At a temperature $T=2$~mK, the resulting theoretical ionization rate
coefficients,
\begin{align}
K_\text{3He} &= \text{$1.7 \times 10^{-10}$~cm$^3$/s,} \\
K_\text{4He} &= \text{$8.0 \times 10^{-11}$~cm$^3$/s,}
\end{align}
agree very well with the experimental values of
Table~\ref{tab5:expres}. At this low temperature only s and p-waves
have to be taken into account as higher order waves contribute
less than a few percent. The samples are partially polarized, so
that quadratic ionization rates are smaller than they would be in
unpolarized samples. Because of the different substate structure
($s=1$ compared to $f=3/2$) and different quantum statistical
symmetry, the corrections for the two isotopes differ,
$K_\text{4He}/K_\text{4He}^\text{(unpol)} = 0.81$ and
$K_\text{3He}/K_\text{3He}^\text{(unpol)} = 0.93$.
For direct comparison with the theoretical results of Table~\ref{tab3:theorycompare} $K^\text{(unpol)}$ for $^3$He* and $^4$He*
have been included in Table~\ref{tab5:expres} as well.

\section{\label{sec:disc}Discussion and conclusions}
The theoretical model presented in Sec.~\ref{sec:th} shows good
agreement with other theoretical work and with the experimental
results reported in Sec.~\ref{sec:exp}. The agreement between the various results has to be considered partly coincidental
in view of the theoretical uncertainties in the potentials, as already mentioned in Sec.~\ref{sec:th:unpol}. Anyway, this shows that cold
ionizing collisions of He* atoms can be understood as single-channel
processes that are determined by Wigner's spin-conservation rule,
quantum threshold behavior and the symmetrization postulate. Using
the model, the difference between the ionization rate coefficients
for $^3$He* and $^4$He* can be interpreted as an effect of the
different quantum statistical symmetry of the two isotopes and the
presence of a nuclear spin in the case of $^3$He*; these differences do not depend on uncertainties related to the potential
as the same potentials (with mass scaling) were used in the $^4$He* and $^3$He* case. As the model is
relatively simple, it is complementary to the more complete (and
precise) close-coupling theory that has been developed for $^4$He*
collisions \cite{vent99,vent00,leo01}. Another single-channel model,
that was published recently \cite{kuma99}, was shown to be based on
erroneous assumptions.

The experimental values of Sec.~\ref{sec:exp} do not agree with
other (mutually inconsistent) experimental results
\cite{mast98,kuma99,tol99,hers00}. The discrepancy between the
experimental values is difficult to interpret, as the experimental
determination of the ionization rate coefficient is the result of an
extensive analysis, including the determination of the density
distribution and temperature of the trapped samples, as well as the
distribution over the magnetic substates. The ionization rate
coefficient is particularly sensitive to the density in the sample.
It must be checked carefully if frequency drifts of the probe laser
light or stray magnetic fields are small enough to avoid incomplete
absorption, which would result in underestimation of the density and
overestimation of the ionization rate coefficients. If the number of
trapped atoms is obtained from fluorescence imaging \cite{kuma99},
an accurate calibration of the \CCD\ chip is crucial. Furthermore,
collisions of trapped atoms with background atoms or atoms from the
atomic beam must be considered. If the trapped atom number is small
($<10^7$), quadratic ionization becomes small and other (linear)
ionization mechanisms possibly play a part, hampering accurate
measurements. Finally, it should be noted that in most experiments
\cite{mast98,kuma99,tol99} the magnetic substate distribution has
not been taken into account.

It would be interesting to extend the work presented here to trapped
samples containing both isotopes and study heteronuclear
ionizing collisions. In the case of collisions between
distinguishable particles, quantum statistical symmetry requirements
should be absent, which could be confirmed from an investigation of
ionizing collisions. This work is in progress in our laboratory.
Another interesting extension of the work
presented here, is the study of the ionization rates for samples
with a prepared substate population. It might be possible to study
depolarization due to collisions \cite{leo01}. Finally, ionizing
collisions can also be investigated in the quantum degenerate regime
\cite{seid04}. Ionization rates could be used to study quantum
statistical properties of a quantum degenerate mixture with high
spatial and temporal resolution. It is conceivable that phase
separation in a quantum degenerate mixture will be observable through
the ionization rate in the sample.

\begin{acknowledgments}
We thank Jacques Bouma for technical support and Elmar Reinhold
for discussions on molecular physics. This work was
supported by the Space Research Organization Netherlands (\textsc{sron}), grant
MG-051, and the European Union, grant HPRN-CT-2000-00125.
\end{acknowledgments}

\bibliography{pra_stas_refs}

\end{document}